\documentclass[prl, twocolumn]{revtex4}%
    \begin{document}
    \title{Bell-type inequalities for partial
    separability in $N$-particle systems and quantum mechanical violations}
    \date{\today}
    \author{Michael Seevinck}
    \email{M.P.Seevinck@student.kun.nl}
    \affiliation{Subfaculty of Physics, University of Nijmegen, \\ Nijmegen, the
    Netherlands  \\ and \\ Institute for History and Foundations of Science,
    University of Utrecht, \\ Utrecht, The Netherlands}
    \author{George  Svetlichny}
    \email{svetlich@mat.puc-rio.br}
    \affiliation{Departamento de Matem\'atica, \\ Pontif{\'{\i}}cia
    Universidade Cat{\'o}lica, \\ Rio de Janeiro, RJ,  Brazil}
                
                
    \begin{abstract}
    We derive $N$-particle Bell-type inequalities under the assumption
    of partial separability, i.e. that the $N$-particle system is composed of
    subsystems which may be correlated in any way (e.g. entangled) but
    which are uncorrelated with respect to each other.  
    These inequalities provide, upon violation, experimentally accessible 
    sufficient conditions for full
    $N$-particle entanglement, i.e. for \(N\)-particle entanglement 
    that cannot be reduced to mixtures of states in which a smaller number of
    particles are entangled.    
    The inequalities are shown to be maximally violated by the $N$-particle
    Greenberger-Horne-Zeilinger (GHZ) states. 
    \end{abstract}
           
    \maketitle
    Given the recent experimental interest in
    quantum correlations in \(3\)- and \(4\)-particle systems described in Refs. 
    \cite{BOU,RAU,SAC,PAN},
    and the probable extension to even larger number of particles,
    it becomes relevant to determine whether such
    correlations are due to full \(N\)-particle quantum entanglement
    and not just classical combinations of quantum entanglement
    of a smaller number of particles.
    We here derive a set of Bell-type inequalities that addresses this question
    by generalizing an analysis of Svetlichny \cite{3bb} from \(3\)-particle to
    \(N\)-particle systems.

    Besides this experimental interest, the Bell-type inequalities here presented
    also
    address the fundamental question of whether Nature somehow limits the number 
    of particles that can be fully entangled, 
    that is to say whether or not some form of partial separability holds.
    Note that this partial separability is distinct from the well studied
    notion of  full separability in Refs. \cite{GISIN,WER,ZHU}.
    In the former the subsets of the \(N\) particles form (possibly entangled) 
    extended systems which however are all uncorrelated with respect to each other, 
    whereas in the latter each particle is
    uncorrelated with respect to all others.
        
    Our results differ from other results on ``\(N\)-particle Bell-type
    inequalities" such as the recent ones found in Refs. \cite{GISIN,WER,ZHU}
    in the following way. None of these treat partial separability in full
    generality. 
    They either assume full separability of all the particles, or full
    separability of some subset, and do not address the issue discussed here.
       
    In this letter the study of partial separability will be 
    shown to result in new types of hidden variable theories and to give
    experimentally accessible conditions that deal with both the experimental and
    the fundamental question mentioned above.

    These experimentally accessible conditions are
    provided by the Bell-type inequalities that we derive from the assumption of
    partial separability. Upon violation they are sufficient conditions for full
    $N$-partite entanglement, i.e. conditions to distinguish between
    $N$-particle states in which all $N$ particles are entangled to
    each other and states in which only $P$ particles are entangled (with
    $P<N$). We also show that these inequalities are maximally violated
    by the $N$-partite  Greenberger-Horne-Zeilinger (GHZ) states. Lastly, we end
    with some concluding remarks that compare our conditions to other similar
    conditions.

    In order to derive our main results, imagine a system decaying into \(N\)
    particles which then separate into \(N\) different directions. At some later
    time we
    perform dichotomous measurements on each of the \(N\) particles,
    represented by observables \(A^{(1)},A^{(2)},\dots A^{(N)}\),
    respectively, with possible results \(\pm 1\). Let us now make the
    following hypothesis of {\it partial separability}: An ensemble of such
    decaying systems consists of subensembles in which each one of the
    subsets of the \(N\) particles form (possibly entangled) extended systems
    which however are uncorrelated to each other.   Let us for the time 
    being focus our attention on one of these subensembles, formed by
    a system consisting of two subsystems of \(P<N\) and \(N-P<N\)
    particles which uncorrelated to each other. Assume also for the time being that
    the
    first subsystem is formed by particles \(1,2,\dots,P\) and the other by the
    remaining. We
    express our partial separability hypothesis by assuming a factorizable
    expression for the probability \( p (a_1a_2\cdots a_N)\) for
    observing the results \(a_i\), for the observables \( A^{(i)}\):
    \begin{eqnarray}\nonumber
    \lefteqn{p (a_1a_2\cdots a_N)=} \\  \label{eq1}
    & &  \int q(a_1\cdots a_P|\lambda)
    r(a_{P+1}\cdots a_N|\lambda)\,d\rho(\lambda),
    \end{eqnarray}
    where \(q\) and \(r\) are probabilities conditioned to the hidden
    variable \(\lambda\) with probability measure \(d\rho\). Formulas
    similar to (\ref{eq1}) with different choices of the composing
    particles and different value of \(P\) describe the other subensembles.
    We need not consider decomposition into more than two subsystems
    as then any two can be considered jointly as parts of one
    subsystem still uncorrelated with respect to the others.   
    Though (\ref{eq1}) expresses a hidden variable model for the local (i.e.
    uncorrelated)  
    behavior of the two subsystems in relation to each other, we shall show that
    the same inequality derived below can be used to distinguish, in the quantum
    mechanical case, between fully entangles states and those only partially
    entangled.

    Consider the expected value of the product of the observables in the
    original ensemble        
    \begin{eqnarray}\nonumber
    \lefteqn{E(A^{(1)}A^{(2)}\cdots A^{(N)})= }  \\ 
    & &   \langle A^{(1)}A^{(2)}\cdots A^{(N)}\rangle =
    \sum_J(-1)^{n(J)}p(J),
    \end{eqnarray}
    where \(J\) stands for an \(N\)-tuple \(j_1,\dots,j_N\) with \(j_k=\pm 1\),
    \(n(J)\) is the number of \(-1\) values in \(J\) and \(p(J)\) is the
    probability of achieving the indicated values of the observables. 
    Using the hypothesis of Eq.\ (\ref{eq1}) as a constraint we now  
    derive non-trivial inequalities satisfied by the numbers
    \(E(A^{(1)}A^{(2)}\cdots A^{(N)})\) when introducing two alternative dichotomous
    observables \(A^{(i)}_1,A^{(i)}_2\), \(i=1,2,\dots ,N\) for each of
    the particles. To simplify the notation we write \(E(i_1i_2\cdots i_N)\) for
    \(E(A^{(1)}_{i_1}A^{(2)}_{i_2}\cdots A^{(N)}_{i_N})\).  For any value of \(P\)
    and any choice of these \(P\) particles to comprise one of the subsystems we
    obtain (proof in the Appendix) the following inequalities:   
    \begin{equation}\label{eq3}
    \sum_I\nu^\pm_{t(I)}E(i_1i_2\cdots i_N)\le 2^{N-1} ,
    \end{equation}
    where \(I=(i_1,i_2,\dots,i_N)\), \(t(I)\) is the number of
    times index \(2\) appears in \(I\), and \(\nu^\pm_k\) is a sequence 
    of signs given by 
    \begin{equation}\label{eq:nupm}
    \nu^{\pm}_k = (-1)^{\frac{k(k\pm 1)}{2}}.
    \end{equation} 
    These sequences have
    period four with cycles \((1,-1,-1,1)\) and \((1,1,-1,-1)\)
    respectively.  We call these inequalities {\em alternating\/}. They are
    direct generalizations of the tri-partite inequalities in Svetlichny \cite{3bb}.
    The
    alternating inequalities are satisfied by a system with {\em any\/} form of
    partial
    separability, so their violation is a sufficient indication of full
    nonseparability.  
              
    Introduce now the operator
    \begin{equation}\label{eq:sop}
    S_N^\pm = \sum_I\nu_{t(I)}^\pm A^{(1)}_{i_1}\cdots A^{(N)}_{i_N}.
    \end{equation}
    Using Eq. (\ref{eq3}) the $N$-particle  alternating
    inequalities can be expressed as 
    \begin{equation}\label{Npart}
    |\langle S^\pm_N\rangle |\le 2^{N-1}.
    \end{equation}
    For even \(N\) the two inequalities are interchanged by a global
    change of labels $1$ and $2$ and are thus equivalent. However for odd \(N\) this
    is not the case and thus they must be considered a-priori independent. To see
    this consider the effect of such a change upon the cycle $(1,-1,-1,1)$. 
    If $N$ is even, we get $(-1)^{N/2}(1 ,1,-1,-1)$ which gives the second
    alternating inequality. For $N$ odd, we get  $\pm(1 ,-1,-1,1)$, which results in
    the same inequality. Similar results hold for the other cycle.

    The two alternating solutions for \(N=2\) are the usual Bell inequalities,
    the  ones for \(N=3\) give rise to the two inequalities found in Svetlichny
    \cite{3bb}\footnote{There is a sign error in front of the \(E(112)\) term of
    equation (6) in \cite{3bb}.}, and for \(N=4\) we have
    \begin{eqnarray*}
    & & \lefteqn{|E(1111)-E(2111)-E(1211)-E(1121)-}\nonumber \\
    & & E(1112)-E(2211)-E(2121)-E(2112)- \nonumber \\
    & & E(1221)-E(1212)-E(1122)+E(2221)+ \nonumber \\
    & & E(2212)+E(2122)+E(1222)+E(2222)|\le 8, 
    \end{eqnarray*}
    where the second inequality is found by interchanging the observable labels $1$
    and $2$. 

    The $N$-particle alternating inequalities were derived for hidden variable
    states $\lambda$. However, they also hold for $N$-partite quantum states which
    are ($N-1$) partite entangled (or non-entangled).  
    In order to see this, suppose we choose the set of all
    hidden variables to be the set of all states on the Hilbert space $\mathcal{H}$
    of the system and $\rho(\lambda)=\delta(\lambda-\lambda_0)$ where $\
    \lambda_0 $ is a quantum state in which one
    particle (say the $N$-th) is independent from the others, i.e.:

    \begin{equation}\label{form}
    \rho = \rho^{\{1,\ldots, N-1\}}\otimes \rho^{\{N\}}.
    \end{equation}
    We then recover the factorizable condition of Eq.(\ref{eq1}):
    \begin{equation}\label{factor}
    p(a_1a_2\cdots a_N|\lambda_0)= p_{\rho^{\{1,\ldots, N-1\}}}
    (a_1a_2\cdots a_{N-1}) p_{\rho^{\{N\}}}(a_N)   
    \end{equation}
    where $p_{\rho^{\{1,\ldots, N-1\}}}
    (a_1a_2\cdots a_{N-1})$ and $p_{\rho^{\{N\}}}(a_N))$ are 
    the corresponding (joint) quantum mechanical probabilities 
    to obtain $a_1,a_2,\cdots, a_N$ for
    measurements of observables $A^{(1)},A^{(2)},\cdots, A^{(N)}$. The expectation
    value $E(A^{(1)}A^{(2)}\cdots A^{(N)})$ then becomes the quantum mechanical
    expression: $E_{\lambda_0}(A^{(1)}A^{(2)}\cdots A^{(N)})=
    {\textrm{Tr}} [\rho^{\{1,\ldots, N-1\}} A^{(1)}
    \otimes A^{(2)}\otimes \cdots \otimes A^{(N-1)}]
    {\textrm{Tr}} [\rho^{\{N\}}A^{(N)}]$.
    Thus the same bound as in the alternating inequalities of Eq.(\ref{Npart})
    holds also for the quantum mechanical expectation values for a state of the
    form Eq.(\ref{form}).

    Since the alternating inequalities of Eq.(\ref{Npart}) are invariant under a
    permutation of the $N$ particles, this bound holds also for a state in which
    another
    particle than the $N$-th factorizes, and, since the inequalities are convex as
    a function of $\rho$, it holds also for mixtures of such states. Hence, for
    every $(N-1)$-particle entangled state $\rho$ we have
    \begin{equation} \label{v}
    |\langle S_N^\pm\rangle_\rho|=|\hbox{Tr}(\rho S_N^\pm)| \leq 2^{N-1}. 
    \end{equation}
    Thus, a  sufficient condition for full $N$-particle entanglement is a violation
    of
    Eq.(\ref{v}).  

    Now from this it follows that using the inequalities of Eq.(\ref{v}) 
    one can experimentally address the fundamental question of whether
    there is a limit to the number of particles that
    can be fully entangled, i.e. whether or not all forms of partial separability
    can be excluded when increasing the number of particles $N$.

    The maximal quantum mechanical violation the left-hand side of the
    \(N\)-particle alternating inequalities of Eq.(\ref{Npart}) is obtained for
    fully entangled $N$-particle  quantum states and is equal to \(2^{N-1}\sqrt2\).
    To see this note that the following recursive relation holds:
    \begin{equation}\label{eq:rec}
    S_N^\pm = S_{N-1}^\pm A^{(N)}_1 \mp S_{N-1}^\mp
    A^{(N)}_2.\end{equation}
    Consider the term \( S_{N-1}^\pm A^{(N)}_1\) which is a self-adjoint
    operator. The maximum \(K\) of the modulus of its quantum expectation
    \(|\langle S_{N-1}^\pm A^{(N)}_1\rangle |\) is equal to the maximum of 
    \(|\langle S_{N-1}^\pm\rangle|\) since the eigenvalues of \(A^{(N)}_1\) are 
    \(\pm 1\). Similarly for the other term. Thus one can take the \(N\)-particle
    bound as twice the \((N-1)\)-particle bound. Since the quantum mechanical bound
    on
    the Bell inequality is \(2\sqrt 2\) the result follows.

    This upper bound is in fact achieved for the
    Greenberger-Horne-Zeilinger (GHZ) states for appropriate values of
    the polarizer angles of the relevant spin observables. Consider the general
    GHZ state:
    \[\Psi_N = \frac{1}{\sqrt 2}( |\!\uparrow\rangle^{\otimes
    N}\pm|\!\downarrow\rangle^{\otimes
    N})=\frac{1}{\sqrt 2}(|\!\uparrow\uparrow\cdots\uparrow\uparrow\rangle \pm
    |\!\downarrow\downarrow\cdots\downarrow\downarrow\rangle ).\] 
    Let 
    \(A^{(k)}_i=\cos\alpha^{(k)}_i \sigma_x +
    \sin\alpha^{(k)}_i \sigma_y\) denote spin observables with angle
    \(\alpha^{(k)}_i\) in the \(x\)-\(y\) plane. A simple calculation
    shows 
    \begin{equation}\label{eq:eghz} E(i_1\cdots
    i_N)=\pm\cos(\alpha^{(1)}_{i_1}+\cdots+\alpha^{(N)}_{i_N}) ,
    \end{equation}
    where the sign is the sign chosen in the GHZ state.

    We now note that for \(k=0,1,2,\dots\) one has:
    \(\cos\left(\pm\frac{\pi}{4}+k\frac{\pi}{2}\right)=\nu^{\pm}_k\frac{\sqrt
    2}{2}\) where \(\nu^{\pm}_k\) is given by (\ref{eq:nupm}). This means that by a
    proper choice of angles, we can match, up to an overall sign, the sign of the
    cosine in Eq.(\ref{eq:eghz}) with the  sign in  front of \(E(i_1\cdots i_N)\) as
    it appears in the inequality, forcing the left-hand side of the inequality to be
    equal to
    \(2^{N-1}\sqrt 2\). This can be easily done if each time an index \(i_j\)
    changes from \(1\) to \(2\), the argument of the cosine is increased by
    \(\frac{\pi}{2}\).  Choose therefore
    \begin{eqnarray*}\left(\alpha_1^{(1)},\alpha_1^{(2)},\dots,\alpha_1^{(N)}\right)
    &=&\left(\pm\frac{\pi}{4},0,\dots,0\right),
    \\ \left(\alpha_2^{(1)},\alpha_2^{(2)},\dots,\alpha_2^{(N)}\right)&=&   
    \left(\pm\frac{\pi}{4}+\frac{\pi}{2},\frac{\pi}{2},\dots,\frac{\pi}{2}\right),
    \end{eqnarray*}
    where the sign indicates which of the two \(\nu^{\pm}\) inequalities is used.

    {\it Concluding remarks}: Recently Seevinck and Uffink \cite{SU} argue that the
    experimental data from some recent experiments (Refs. \cite{BOU,RAU,PAN})
    designed to produce full three particle entangled states do  not completely rule
    out the hypothesis of a partially entangled state. Further, an analysis of these
    experiments shows
    that the three particle alternating inequalities presented above would not be
    violated by the choice of the experimental observables and thus, based on the
    present
    inequalities, full entanglement in these experiments is still not
    established. However, we hope that future experiments (including
    $N=4$  and higher) will yield experimental tests of the alternating
    inequalities and will thereby provide conclusive tests for the existence of
    full $N$-partite entanglement. See also Cereceda \cite{CER} for another analysis
    of this point.

    Similar conditions as the ones presented here to test for full
    entanglement were obtained in Gisin and Bechmann-Pasquinucci \cite{GISIN}. 
    However these differ in at least two aspects. 
    Firstly, for $N$ even they are equivalent whereas for $N$ odd this is not the
    case    
    (see Uffink \cite{UFFINK}). Secondly and more importantly, the inequalities of
    Ref. \cite{GISIN} are only derived quantum mechanically, i.e they only hold for
    quantum systems, and are thus unable to address the general requirement 
    of partial separability which has been treated here.

    After the present paper was submitted for publication, an article by Collins
    et.\ al.\ \cite{COL} appeared that treats some of the same questions and gives
    an independent proof of our alternating inequalities. 
             
    \section{Appendix: Proof of inequality (\ref{eq3})}
    We seek inequalities of the form
    \begin{equation}
    \sum_I\sigma_IE(i_1i_2\cdots i_N) \le M ,    \end{equation}
    where \(\sigma_I\) is a sign and $M$ nontrivial.
    Following almost verbatim the analysis in \cite{3bb}, one must look for
    \(\sigma_I\) which solve the minimax problem
    \begin{equation}\label{eq:minmax}
    m=\min_\sigma m_\sigma =\min_\sigma\max_{\xi,\eta}
    \sum_{I}\sigma_I\xi_{i_1\cdots i_P}\eta_{i_{P+1}\cdots i_N},
    \end{equation}
    where \(\xi_{i_1\cdots i_P}=\pm 1\) and \(\eta_{i_{P+1}\cdots i_N}=\pm 1\)
    are also signs.  Without loss of generality we can take  \(P\ge N-P\).

    One can derive some useful
    upper bound on \(m\). Toward this end, set $\eta_{i_{P+1}\cdots i_{N-1}2}=
    \zeta_{i_{P+1}\cdots i_{N-1}}\eta_{i_{P+1}\cdots i_{N-1}1}$ for some sign
    \(\zeta_{i_{P+1}\cdots i_{N-1}}\), using the fact that $i_N=1,2$. Taking into
    account that \(\sigma_I^2=1\), and denoting by \(\hat I\) the \((N-1)\)-tuple
    \((i_1,\dots,i_{N-1})\) we have:
    \begin{eqnarray*}
    \lefteqn{m_\sigma =} \\
    & & \max \sum_{\hat I} \sigma_{\hat I1}\eta_{i_{P+1}\cdots i_{N-1}1}
    \xi_{i_1\cdots i_P}(1  + \sigma_{\hat I1}\sigma_{\hat
    I2}\zeta_{i_{P+1}\cdots
    i_{N-1}}).
    \end{eqnarray*}
    The maximum being over  \(\xi_{i_1\cdots i_P}\), \(\eta_{i_{P+1}\cdots
    i_{N-1}1}\), and \(\zeta_{i_{P+1}\cdots i_{N-1}}\).

    Now certainly one has
    \begin{equation}\label{eq:rm}
    m_\sigma \le \hat m_\sigma=\max\sum_{\hat I}|1  + \sigma_{\hat
    I1}\sigma_{\hat I2}\zeta_{i_{P+1}\cdots i_{N-1}}|,
    \end{equation}
    the maximum taken over \(\zeta_{i_{P+1}\cdots i_{N-1}}\).

    If we define \( \hat m=\min_\sigma\hat m_\sigma\) one easily sees that
    \(\hat m = 2^{N-1}\). This can only be achieved under the following condition:
    \begin{equation}\label{eq:mincon} \begin{array}{l} \hbox{\sl For each fixed\ }
    (i_{P+1},\dots, i_{N-1}) \hbox{\sl\ exactly\ } 2^{P-1} \\
    \hbox{\sl of the quantities\ } \sigma_{\hat I1}\sigma_{\hat I2} \hbox{\sl\
    are\ } +1 \hbox{\sl\ and\ } 2^{P-1} \hbox{\sl\ are\ } -1.\end{array}
    \end{equation}
    Although it may be that \(m<\hat m\) we have proven that \(m=\hat m=8\)
    in all cases for \(N=4\), and \(m = \hat m\) for \(P=N-1\) for any \(N\).

    We shall call any choice of the \(\sigma_I\) satisfying this condition
    a {\em minimal solution\/}.

    What immediately follows from the above is that any
    solution of (\ref{eq:mincon}) for a given value of \(P\) is a
    solution for all greater values of \(P\le N-1\).
    A violation of an inequality so obtained for the smallest possible
    value of \(P\ge N/2\) precludes then {\em any\/} factorizable
    model of the \(N\)-particle correlations.

    Assume provisionally that the only decays are those in which an
    \(N=P+(N-P)\) factorization occurs. The whole ensemble of decays
    consists of subensembles corresponding to different choices of the
    \(P\) particles.  We do not know in any particular instance of decay
    to which of the subensembles the event belongs. 
    To take account of this, our inequality must be one that would arise under any
    choice of the \(P\) particles.  Call a minimal solution \(\sigma_I\) 
    {\em admissible\/} if \(\sigma_{\pi(I)}\) is also a minimal
    solution for any permutation \(\pi\). An inequality that follows
    from an admissible solution will therefore be one that must be
    satisfied by any subensemble of \(N=P+(N-P)\) factorizable events.

    The set of admissible solutions  breaks up into orbits by the
    action of the permutation group. The overall sign of \(\sigma_I\)
    is not significant and two solutions that differ by a sign are
    considered equivalent. The set of these equivalence classes also
    breaks up into orbits by the action of the permutation group. It
    is remarkable that there are orbits consisting of one equivalence
    class only. For such, one must have \(\sigma_{\pi(I)} = \pm
    \sigma_{I}\). The sign in front of the right-hand side must be a
    one-dimensional representation of the permutation group, so one
    must have either \(\sigma_{\pi(I)} =  \sigma_{I}\) or
    \(\sigma_{\pi(I)} =  (-1)^{s(\pi)}\sigma_{I}\), where \(s(\pi)\)
    is the parity of \(\pi\). The second case is impossible since one
    then would have \(\sigma_{11\cdots 1}=-\sigma_{11\cdots 1}\) as a
    result of a flip permutation. Since an overall sign is not
    significant one can now fix \(\sigma_{11\cdots 1}=1\).  As the
    only permutation invariant of \(I\) is \(t(I)\), the number of
    times index \(2\) appears in \(I\), we must have \(\sigma_I =
    \nu_{t(I)}\) for some \((N+1)\)-tuple (\(\nu_0=1\) by convention)
    \(\nu=(1,\nu_1,\nu_2,\dots,\nu_N)\). We must now solve for the
    possible values of \(\nu\).

    Let \(a=t(i_{P+1}\cdots i_{N-1})\) and \(b=t(i_1\cdots i_P)\), then
    condition  (\ref{eq:mincon}) for our choice of \(\sigma_I\), is equivalent
    to
    \(\nu\) satisfying
    \begin{equation}\label{eq:nucond}
    \sum_{b=0}^P\left(\begin{array}{c}P \\ b\end{array}\right)
    \nu_{a+b}\nu_{a+b+1} = 0,\quad a=0,1,\dots N-P-1.
    \end{equation}
    Let $\mu_k=\nu_k\nu_{k+1}$. Eq. (\ref{eq:nucond}) then becomes
    \begin{equation}\label{eq:mucond}
    \sum_{b=0}^P\left(\begin{array}{c}P \\ b\end{array}\right) \mu_{a+b} =
    0,\quad a=0,1,\dots N-P-1.
    \end{equation}

    Now it is obvious that there are at least two solutions of (\ref{eq:mucond})
    valid for all \(P\), to wit \(\mu_k = \pm (-1)^{k}\) since then
    (\ref{eq:mucond}) is just the
    expansion of \((1-1)^P\) or \((-1+1)^P\). Call these solutions the {\em
    alternating\/} solutions.  Finally we get from $\mu_k=\nu_k\nu_{k+1}$ the two
    solutions (\ref{eq:nupm}) once we've chosen the overall sign to set 
    \(\nu_0=1\).

    \subsection*{Acknowledgements}
    We are very grateful to Jos Uffink for stimulating
    discussions, critique and valuable ideas. One of us (GS) thanks the Conselho
    Nacional de Desenvolvimento Cient\'{\i}fico e Tecnol\'ogico (CNPq) for financial
    support.

    \end{document}